\documentclass[a4paper,11pt]{article}
\pdfoutput=1
\usepackage{jcappub}

\usepackage{amsmath}
\usepackage{bm}
\usepackage{xcolor,color,colortbl}
\usepackage[utf8]{inputenc}
\usepackage{hyperref}
\usepackage{url}
\usepackage{graphicx}
\usepackage{multirow,bigdelim}
\usepackage{amssymb}
\usepackage[amssymb]{SIunits}
\usepackage{multirow}
\usepackage{bbold}
\usepackage{verbatim}
\usepackage{comment}
\usepackage{caption} 
\usepackage{physics}
\usepackage{subcaption}
\usepackage{orcidlink}

\usepackage{tikz}
\usetikzlibrary{positioning}

\definecolor{darkgreen}{HTML}{339933}

\usepackage{scalerel,tikz}
\usetikzlibrary{svg.path}
\definecolor{orcidlogocol}{HTML}{A6CE39}
\tikzset{orcidlogo/.pic={
 \fill[orcidlogocol] svg{M256,128c0,70.7-57.3,128-128,128C57.3,256,0,198.7,0,128C0,57.3,57.3,0,128,0C198.7,0,256,57.3,256,128z};
 \fill[white] svg{M86.3,186.2H70.9V79.1h15.4v48.4V186.2z}
 svg{M108.9,79.1h41.6c39.6,0,57,28.3,57,53.6c0,27.5-21.5,53.6-56.8,53.6h-41.8V79.1z M124.3,172.4h24.5c34.9,0,42.9-26.5,42.9-39.7c0-21.5-13.7-39.7-43.7-39.7h-23.7V172.4z}
 svg{M88.7,56.8c0,5.5-4.5,10.1-10.1,10.1c-5.6,0-10.1-4.6-10.1-10.1c0-5.6,4.5-10.1,10.1-10.1C84.2,46.7,88.7,51.3,88.7,56.8z};
}}
\newcommand\orcidicon[1]{\href{https://orcid.org/#1}{\mbox{\scalerel*{
\begin{tikzpicture}[yscale=-1,transform shape]
\pic{orcidlogo};
\end{tikzpicture}
}{|}}}}

\title{Probing baryonic feedback using fast radio bursts and galaxy surveys}

\author[a,b,c]{Haruki Ebina~\orcidicon{0000-0002-1080-0955}}
\author[a,b,c]{and Martin White~\orcidicon{0000-0001-9912-5070}}
\affiliation[a]{Department of Physics, University of California, Berkeley, CA 94720, USA}
\affiliation[b]{Berkeley Center for Cosmological Physics, UC Berkeley, CA 94720, USA}
\affiliation[c]{Lawrence Berkeley National Laboratory, One Cyclotron Road, Berkeley, CA 94720, USA}

\emailAdd{ebina@berkeley.edu}

\abstract{
The dispersion measure (DM) of fast radio bursts (FRB) provides a noisy measure of the free-electron column density to the burst. With next-generation FRB surveys the number of bursts will drastically increase and measurements of the baryon distribution will be possible on the Mpc scales most relevant to Stage IV weak lensing surveys.  Comparing the DM auto-spectrum, DM-shear cross-spectrum and the kinetic Sunyaev-Zel'dovich effect, we argue that the cross-correlation of FRB catalogs with galaxy surveys offers many advantages.  Hybrid Effective Field Theory (HEFT), often used in large-scale structure cosmology, is capable of modeling the clustering of electrons jointly with galaxies and matter to small scales.  Using HEFT with the Pad\'e model of baryonic feedback, we forecast that next-generation FRB surveys (combined with galaxy samples) will be able to constrain feedback at a level informative for Stage~IV weak lensing surveys. In particular we should not be limited by modeling the manner in which galaxies trace the matter field.  Rather constraints of the higher-order electron biases and the electron-galaxy cross-stochasticity provide the dominant limitation, and are an obvious target for improvement in order to reach percent-level constraints.  The same measurements also provide a valuable means of constraining feedback in hydrodynamic simulations of galaxy formation in a way that is not tied to the modeling of particular galaxy samples.
}

\begin{document}
\maketitle
\flushbottom

\section{Introduction}

We are entering the Stage IV era of cosmic shear surveys, with ESA's Euclid satellite \cite{Euclid}, the Legacy Survey of Space and Time (LSST) on the Vera C Rubin observatory \cite{LSST} and the Nancy Grace Roman space telescope \cite{Roman} all set to return dramatically improved datasets for the study of weak lensing with galaxies.  With their dramatic increase in statistical precision, these surveys also become sensitive to small systematic effects.  One such effect is the impact of feedback and baryonic physics on the matter power spectrum, which under many plausible scenarios can seriously bias cosmological constraints from Stage IV surveys if not properly accounted for \cite{Mandelbaum18,Chisari19,Amon22}.  
This effect will be even more pronounced for kinematic lensing measurements, which constrain the same signal with nearly an order of magnitude smaller shape noise \cite{Blain02,Huff13,Pranjal23}. 
The ideal situation would be to include a flexible model for the impact of baryons, to avoid bias, that is constrained by external data, to avoid a dramatic increase in the uncertainties on cosmological parameters.  While flexible models for the shape of this feedback exist, the translation between small-scale data usually used to constrain them and the baryonic suppression on the intermediate scales relevant for cosmology with cosmic shear can be difficult and model dependent.

Both the kinematic Sunyaev-Zeldovich (kSZ; \cite{kSZ}) effect and the dispersion measure (DM) of Fast Radio Bursts (FRBs; \cite{Bhandari21,Petroff22}) have been proposed as providing external constraints on the distribution of baryonic material.  Both probes involve the distribution of ionized gas, i.e.\ the electron density.  At $z\lesssim3$, free electrons reside overwhelmingly in the diffuse ionized phase that hosts nearly 90\% of the baryons at low redshift \cite{McQuinn14,Peroux20} and trace the underlying baryon field at the few-percent level \cite{Masui15,Leung25,Zhou25b,Andrew26,Ebina26d,Sullivan26}.
Our focus here will be on the FRBs, where we expect a rapid increase in observational capabilities within the next few years.  We shall comment upon kSZ later.  While current-generation surveys \cite{CHIME,Lanman24,askap,Law24} are capable of detecting $\mathcal{O}(100)$ localized FRBs per year \cite{Lanman24,JahnsSchindler23}, next-generation surveys (like the Deep Synoptic Array \cite{Hallinan19,Connor26}, Canadian Hydrogen Observatory and Radio-transient Detector \cite{CHORD} and Square Kilometre Array \cite{Macquart15,Zhang23}) are expected to detect $\mathcal{O}(10^4)$ localized FRBs per year \cite{Hallinan19,CHORD,Connor26}.  Taking the detection capability of next-generation experiments into account, the redshift distribution of detected FRBs is likely to peak around $z\sim0.6$ with the high-$z$ distribution tail extending past $z\sim2$ \cite{Feng26}.  The exact shape is highly uncertain, but where necessary we will take $dN/dz\propto z^2\exp[-z/z_0]$ with $z_0=0.2$ so that $\langle z\rangle=3\,z_0=0.6$ and the median $z_{\rm med}\simeq 2.67406\,z_0\simeq 0.535$.

An advantage of FRB DMs is that they probe the gas distribution on the scales directly relevant to cosmic shear surveys and that the observables can be naturally cast in the language familiar to large-scale structure.  This allows us to immediately reuse many of the tools developed by the community in the last few decades. The free electrons measured through FRB DMs can be modeled as a (weakly) biased tracer of the underlying matter, with a linear bias close to unity and very small higher-order biases \cite{Masui15,Zhou25b,Andrew26,Ebina26d,Sullivan26}.  The dispersion measure provides a noisy integral of the electron density, and thus the gas density along the line of sight, to each FRB.  This view of FRBs provides the advantage that the formalism and many of the parameters are shared between the cosmic shear analysis and the FRB `prior' that is being used to constrain the feedback in the shear analysis.

The purpose of this paper is to provide a preliminary exploration of how and how well FRBs can be used to provide priors on baryonic feedback for Stage IV cosmic shear surveys.  The work is organized as follows. First, in \S\ref{sec:frbintro}, we introduce the FRB dispersion measures as a large-scale structure probe, and seek insight into the requirements on the FRB catalogs, galaxy samples, and theoretical modeling for next-generation experiments. Next in \S\ref{sec:theory}, we describe the formalism to model the FRB dispersion auto-correlation and dispersion-galaxy cross-correlation on the same footing as galaxy surveys.  Section \ref{sec:forecasting} describes our Fisher forecasting methodology for translating observational constraints on FRB DMs into priors on the impact of baryonic feedback on $P_{mm}$ while \S\ref{sec:results} gives such forecasts under a variety of different assumptions.
Finally, we conclude in \S\ref{sec:conclusion}.  A number of technical details are relegated to the appendices.

\section{Fast Radio Burst Dispersion Measure}
\label{sec:frbintro}

Each FRB provides an integrated measure of the line-of-sight density of free electrons through the frequency-dependent delays in the observed pulse.  Each FRB provides a (noisy) measurement of the `dispersion measure'\footnote{As $D\propto \Delta t_{\rm obs}\,\nu_{\rm obs}^2$, time dilation and frequency redshifting require $D\propto n_e\,d\ell/(1+z)$ with proper lengths and densities.  Converting to comoving distances and densities gives Eq.~\ref{eq:D}.} 
\begin{equation}
    D(\hat{\mathbf{n}}) =  \int_0^{\chi_f} d\chi \  \bar{n}_e^{\rm com}(z)\,(1+z)(1 + \delta_e(\hat{\mathbf{n}}, z)) = \int_0^{\chi_f} d\chi \, W_D(\chi)(1 + \delta_e(\hat{\mathbf{n}}, z))
    \label{eq:D}
\end{equation}
where $\hat{\mathbf{n}}$ is the direction of the line of sight, $\chi$ is the comoving distance, $\chi_f$ is the comoving distance to the FRB, $z$ is the redshift corresponding to $\chi$, $\delta_e(\hat{\mathbf{n}}, z)$ is the overdensity of free electrons at redshift $z$, and $\bar{n}_e^{\rm com}(z)$ is the comoving, mean free electron density. In the second equality we introduce the DM kernel $W_D$. Note we use the comoving density here, while it is often common to find the physical density instead -- these differ by $(1+z)^3$.

As DM ($D$) is an integrated measure, one cannot measure the 3D clustering directly. Instead, one can analyze the two-point correlation of these fields, e.g.\ through their angular power spectrum.  Assuming bursts at a single redshift (see Appendix \ref{sec:zweight}) and applying the Limber approximation \cite{Limber,LovAfs08,Wayland26,Feng26}
\begin{equation}
    C_\ell^{DD} = \bar{n}_{e,0}^2 \int_0^{\chi_f} d\chi \, \frac{(1+z)^2}{\chi^2} \, \left(\frac{f_e(z)}{f_{e,0}}\right)^2 P_{ee}\left(k=\frac{\ell+1/2}{\chi}, z\right)
\label{eq:ClDD}
\end{equation}
where $\bar{n}_{e,0}=\bar{n}_e^{\rm com}(z=0)$ and we have defined $f_e(z)$ to be the fraction of baryons residing in the diffuse ionized gas at redshift $z$ with $f_{e,0}=f_e(z=0)$.  At $z<1$ we have $f_e(z)/f_{e,0}\approx 1$ \cite{Leung25,Zhou25b,Andrew26} and $\bar{n}_e^{\rm com}(z)$ is independent of redshift.
To compute $C_\ell^{DD}$ we take $P_{ee}$ to be Hybrid Effective Field Theory (HEFT) fits to the FLAMINGO simulation \cite{Schaye23,Kugel23,Helly26}, with higher-order biases and shot noise set to zero. The HEFT spectra are used up to $k=3.5\,h\,\mathrm{Mpc}^{-1}$ and then extrapolated further using \texttt{HaloFit} \cite{Takahashi12}. An additional Gaussian damping of $\Lambda=10\,h\,\mathrm{Mpc}^{-1}$ is applied. The Gaussian damping accounts for our lack of knowledge at high $k$ and will affect $C_\ell^{DD}$ at $\sim10\%$ at $\ell\approx1000$; we believe that for this work this impact is limited, due to noise domination above $\ell\approx400$ as shown in Fig.~\ref{fig:FRBz}, but we discuss this further in Appendix \ref{sec:Pee}.  The generalization of Eq.~\ref{eq:ClDD} to a distribution of FRB redshifts can be found in Appendix \ref{sec:zweight} or refs.~\cite{Wayland26,Feng26}.

Note that the integral in Eq.~\ref{eq:ClDD} contains a lot of weight at very low $\chi$, so that $C_\ell^{DD}$ at even relatively low $\ell$ can contain contributions from $P_{ee}$ at high $k$, which motivates the conservative Gaussian damping above.  The sensitivity to high $k$ has been noted before \cite{McQuinn14} as a way to constrain the small-scale electron density perturbations.  For modeling larger scales it can be handled using a generalization of the `lensing counter terms' idea \cite{DeRose26b}, as also described in Appendix \ref{sec:Pee}.  

The observed DM field contains both the cosmological signal, $D$, and noise with the latter receiving contributions from measurement error, the stochastic host-galaxy and circumburst plasma and residual errors in the Milky Way subtraction.  If this noise is assumed uncorrelated from burst to burst its power spectrum is white and for a large number of sources randomly distributed on the sky
\begin{equation}
  N_\ell^{DD} = \frac{\sigma_{D}^2}{\bar{n}_{\rm FRB}},
\label{eq:NellDD}
\end{equation}
where $\bar{n}_{\rm FRB}$ is the angular number density of FRBs and $\sigma_D$ is the per-burst rms of $\epsilon$. For cosmological FRBs, $\epsilon$ is expected to be dominated by the host and circumburst contribution. Recent fits to localized samples motivate $\sigma_{D} \approx 100\,\mathrm{pc\,cm^{-3}}$ \cite{Connor25}, and this value has been adopted in recent forecasts \cite{Sharma26,Andrew26}, although the inferred scatter carries sizeable uncertainties and depends on the FRB sample. We shall fix $\sigma_{D}=100\,\mathrm{pc\,cm^{-3}}$ and interpret $\bar{n}_{\mathrm{FRB}}$ in Eq.~\ref{eq:NellDD} as an `effective' number density (as is commonly done in cosmic shear studies).  If $\sigma_D^2$ evolves with redshift, this simply modulates the redshift dependence of $\bar{n}_{\rm eff}$.

\begin{figure}
    \centering
    \includegraphics[width=\linewidth]{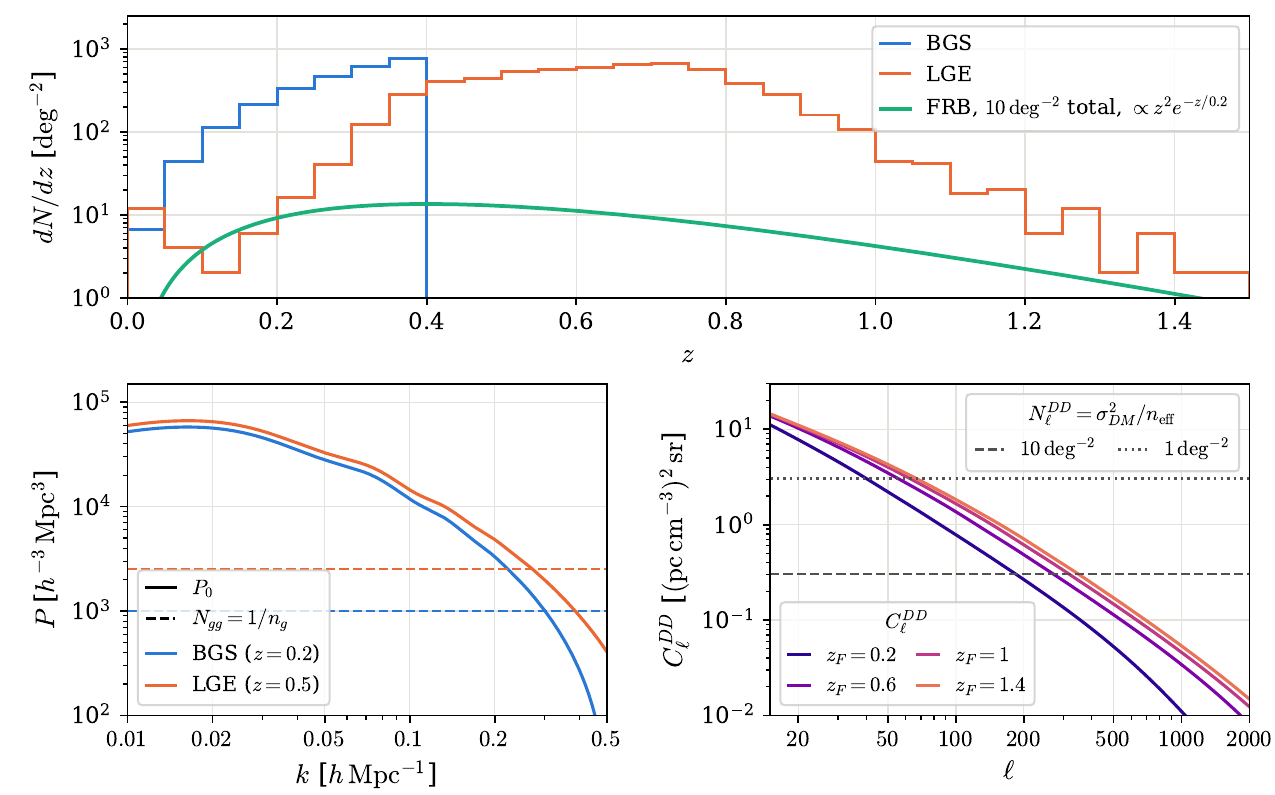}
    \caption{The landscape of FRBs and galaxies (with spectroscopic redshifts) expected in the coming years. Top: The redshift distribution of DESI galaxies (BGS from Run 1, LGE from Run 1b) against that expected from FRBs with $dN/dz\propto z^2e^{-z/z_0}$, for $z_0=0.2$, with a total density of $10\deg^{-2}$. Lower left: The signal to noise of the power spectrum monopole $P_0$ against the Poisson shot noise anticipated from BGS and LGE, at $z=0.2$ and 0.5, respectively.  Lower right: The angular power spectra $C_\ell^{DD}$ of FRB dispersion assuming one source redshift for the FRB population, $z_F$, compared against the shot noise $N_\ell^{DD}$ at effective densities of 1 and $10\deg^{-2}$ with $\sigma_{DM}=100\,\mathrm{pc}\,\mathrm{cm}^{-3}$.}
    \label{fig:FRBz}
\end{figure}

We shall gain redshift information by cross-correlating the DM field with galaxies.  This will also serve to tame the high-$k$ tail of the integral in Eq.~\ref{eq:ClDD}.  Given a galaxy population with a narrow redshift distribution around redshift $z_g$, the cross-spectrum is
\begin{equation}
    C_\ell^{Dg} = \bar{n}_{e}^{\rm com}(z_g) \, \frac{1 + z_g}{\chi_g^2} \, P_{ge}\left(k=\frac{\ell+1/2}{\chi_g}, z_g\right)
\end{equation}
where $\chi_g$ is the corresponding comoving distance and $P_{ge}(k, z_g)$ is the three-dimensional cross-spectrum between the galaxy overdensity field, $\delta_g$, and the free electron overdensity field, $\delta_e$, evaluated at redshift $z_g$ \cite{Shirasaki17,Madhavacheril19}. We refer the reader to refs.~\cite{Rafiei-Ravandi20,Alonso21,Sharma26,Wayland26} for expressions and derivations of $C_\ell^{Dg}$ involving a general (not thin-shell) population of galaxies and FRBs (see also Appendix \ref{sec:zweight}).  Although we shall not make use of it in this paper, we also note that cross-correlation of the FRBs with a well-filled and wide-area galaxy survey strongly reduces sidelobes in the mode coupling matrix relating the measured $C_\ell$ to the theory \cite{Hivon02}.  This will be important until FRB surveys start to probe many sightlines across the sky.

Current or near-future spectroscopic samples provide both the redshift coverage (Fig.~\ref{fig:FRBz}) and the number density required for these cross-correlations.  In all scenarios, the angular number density of galaxies far exceeds that of the FRBs and the uncertainty in the cross-correlation is dominated by the electron noise rather than the galaxy noise, even if we restrict to populations with spectroscopic redshifts.  The gain from using photometric samples is thus due primarily to their footprint \cite{Bilicki14,Peacock16,Schlafly19,Flewelling20,Zhou23,Chen24b,Bock26}, rather than to the enhanced number density.  The lower-left panel of Fig.~\ref{fig:FRBz} shows that both the BGS and LGE populations are sample-variance limited on the scales usually used to model 3D galaxy clustering, so we expect tight constraints on the manner in which galaxies trace the dark matter field (i.e.\ bias).  At $z=0.2$ the mode $k=0.3\,h\,\mathrm{Mpc}^{-1}$, where $\bar{n}P_0\simeq 1$, has $\ell\approx k\chi\approx 200$.  This is noise dominated for FRBs with $\bar{n}_{\rm eff}(z_F>0.3)\lesssim 10\,\mathrm{deg}^{-2}$.  At $z=0.5$ the same mode has $\ell\approx k\chi\approx 400$, and the FRBs are again noise dominated for densities $\lesssim 10\,\mathrm{deg}^{-2}$.  Since $dN/dz$ of the FRBs peaks below $z=0.5$, achieving the same density of FRBs above $z=0.65$ requires more FRBs overall.

The lower-right panel of Fig.~\ref{fig:FRBz} shows $C_\ell^{DD}$ for a population of FRBs all at $z_F$, for $z_F$ ranging from $0.2$ to $1.4$.  For comparison the noise for $\bar{n}_{\rm eff}=1$ and $10\,\mathrm{deg}^{-2}$ are shown as horizontal lines, assuming all of those FRBs are behind the redshift slice of interest.  Note that the DD auto-power integral at these $\ell$s is dominated by relatively low redshifts, so that it converges rapidly as $z_F$ is increased. We refer the reader to Appendix \ref{sec:Pee} for further discussion on the information content of $C_\ell^{DD}$.

\section{Theoretical model}
\label{sec:theory}

In order to interpret the FRB and galaxy observables ($C_\ell^{DD}$, $C_\ell^{Dg}$, $P_{\ell}^{gg}$) we need a model capable of simultaneously predicting $P_{ee}$, $P_{eg}$ and  $P_{gg}$; ideally with a common set of nuisance parameters.  The modeling of the galaxy auto-power is most commonly done within cosmological perturbation theory\footnote{We shall use the \texttt{velocileptors} code \cite{Chen20a,Chen20c} in its Lagrangian formulation, to match the basis used in HEFT.  This code has also been recently used to model galaxy clustering, CMB lensing and cosmic shear \cite{DESI24-V,Chen24b,Maus25b}.}, using the bias expansion (see e.g.\ refs \cite{Desjacques18,Ivanov22b} for recent reviews).
We follow a similar procedure for the electron overdensity field, $\delta_e$.  The concept of modeling the electrons as a linearly biased field of the matter density field ($\delta_e=b_e\delta_m$) has been suggested in the past (e.g.~\cite{Masui15}). 
Recently, refs.~\cite{Zhou25b,Andrew26,Ebina26d} confirmed $b_e\approx 1$ using hydrodynamical simulations.
Refs~\cite{Ebina26d,Sullivan26} generalized this treatment to higher orders and showed that the scale-dependent bias terms are small in the current generation of simulations.  Ref.~\cite{Ebina26d} also introduced Hybrid Effective Field Theory (HEFT; \cite{Modi20,Kokron21,Hadzhiyska21,Zennaro22,PellejeroIbanez23,DeRose23b,Zennaro23,Nicola24,Zhou25a,Bartlett26,PellejeroIbanez26}) as a natural formalism for modeling $P_{ee}$ and $P_{ge}$ (and hence $C_\ell^{DD}$ and $C_\ell^{Dg}$), validating the model against simulations in the process.  This allows us to model the FRB-galaxy cross-clustering at higher precision and to significantly smaller scales while introducing the key physics that leads to decorrelation between the observed cross-spectra and the desired auto-spectra and hence limits our ability to use FRBs to constrain baryonic feedback.  We shall follow this approach in our forecasts.

In HEFT, one employs N-body simulations for the Lagrangian displacement, $\boldsymbol{\Psi}$, while retaining the general and robust perturbative framework of the bias expansion.  We shall work throughout to quadratic order in the bias expansion; extensions beyond quadratic order can be found in refs~\cite{Shiferaw25,Zhou25a,Bartlett26}.
The tracer density field (matter, electrons or galaxies) is then a linear combination of `advected' (time-evolved) fields $F(\mathbf{k})$
\begin{equation}
    \delta_t(\mathbf{k}, a) = \left( 1 - \frac{b_{\nabla} k^2}{4} \right) \delta_{cb}(\mathbf{k}) + \sum_i b_i F_i(\mathbf{k}) + \mathcal{E}(\mathbf{k})
\label{eq:delta_t}
\end{equation}
where $i\in\{\delta,\delta^2,s\}$ and $\mathcal{E}$ represents stochastic noise.  The case where all of the $b_X=0$ corresponds to the matter field.

This can directly be used to predict the auto- and cross-spectrum between tracers.
\begin{equation}
    P^{AB}(k) = P_{11} + \sum_i (b_i^A+b_i^B) P_{1i} + \sum_{ij} b_i^Ab_j^B P_{ij}
    - \frac{1}{4}\left(b_\nabla^A+b_\nabla^B\right) k^2 P_{\rm lin}
    + \cdots
\label{eq:P_HEFT}
\end{equation}
with the component spectra $P_{ij}(k)$ obtained directly from N-body simulations.  The first term, $P_{11}$, is the matter power spectrum.

We make one change to the `standard' HEFT formalism in order to deal with baryonic feedback.  The standard Jeans argument implies that on large scales the impact of baryonic feedback on the matter power spectrum can be captured by the $b_\nabla$ terms in Eq.~\ref{eq:delta_t} \cite{Chen24b,Bartlett26,DeRose26c}.  For the scales probed by the galaxy auto-spectrum this will be sufficient.  However if we wish to extend the model to $k\approx 1\,h\,\mathrm{Mpc}^{-1}$, so as to model cosmic shear, then we need to go beyond this lowest-order response.  A simple extension that fits all of the current generation of hydrodynamic simulations is the Pad\`e model, which replaces $k^2P_{\rm lin}$ in  Eq.~\ref{eq:P_HEFT} with $k^2P_{\rm lin}/[1+(kR)^2]$ where $R$ is a new parameter \cite{Chen24b,Bartlett26,DeRose26c}. 
In general, one would expect $R$ to depend on the tracer population. 

\begin{figure}
    \centering
    \includegraphics[width=\linewidth]{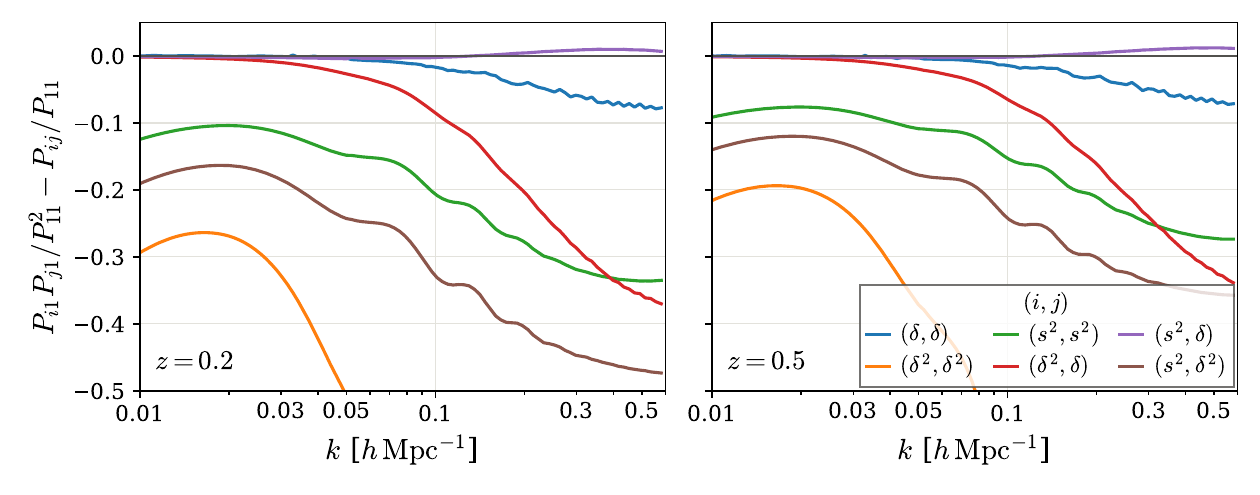}
    \caption{The lowest order (in the bias expansion) contribution to the electron-matter correlation coefficient $r_{em}$ (Eq.~\ref{eq:rem_taylor}), calculated using HEFT. These factors allow us to convert $r_{em}\approx 1$ constraints to limits on biases. 
    }
    \label{fig:heft_decorr}
\end{figure}

With the expressions for the cross- and auto-spectra in Eq.~\ref{eq:P_HEFT} we can also compute the electron-matter cross-correlation coefficient.  For both fields the stochastic noise is negligible and so the deviation of $r_{em}$ from unity is driven primarily by non-linear clustering and non-linear bias.  Specifically
\begin{align}
    r_{em} &\simeq 1 - \frac{1}{2} \sum_{ij} b_{i,e}b_{j,e}
    \left(   \frac{P_{ij}}{P_{11}} - \frac{P_{i1} P_{j1}}{P_{11}^2} \right) + \cdots \label{eq:rem_taylor} \\
    &\approx 1 - \frac{1}{2}b_{1,e}^2
    \underbrace{\left(\frac{P_{\delta\delta}}{P_{11}}-\frac{P_{1\delta}^2}{P_{11}^2}\right)}_{\to 0\mathrm{\ as\ }k\to0} + \cdots
    \label{eq:rem_b1}
\end{align}
where in the second line we have assumed that the $b_{n>1,e}$ terms are small compared to $b_{1,e}$.  This is true in hydrodynamic simulations \cite{Ebina26d,Sullivan26}.  At low $k$ we have $P_{11}\approx P_{1\delta}\approx P_{\delta\delta}\approx P_{\rm lin}$ so the correlation is expected to approach $1$, as observed.  We plot $P_{i1}P_{j1}/P_{11}^2-P_{ij}/P_{11}$ in Fig.~\ref{fig:heft_decorr}, where it is easy to see that by $k\simeq 0.3\,h\,\mathrm{Mpc}^{-1}$ it reaches $\mathcal{O}(0.1-1)$ for almost all terms.  Requiring $r_{em}$ be close to unity thus provides strong limits on the $b_{i,e}$ (see also ref.~\cite{Sullivan26}).

Note that we work throughout in terms of Lagrangian biases.  These can be converted directly to the more familiar Eulerian biases.  The Eulerian bias expansion would model the electron density field $\delta_e$ as a function of the matter field at the time of observation
\begin{equation}
    \delta_e = b_{e,1}^E \delta + \frac{1}{2}b_{e,2}^E \delta^2 + b_{e,s}^E s^2 + \cdots
\end{equation}
as opposed to initial fields in the Lagrangian framework.
Accounting for the evolution from initial to final times one finds\footnote{Some authors use $\mathcal{G}_2$ instead of $s^2$, for which $b_2^E=2b_2+(8/3)b_s$, $b_{\mathcal{G}_2}^E=b_s-(2/7)b_1$ \cite{Chen26}.} (e.g.~refs~\cite{Desjacques18,Chen26})
\begin{equation}
  b_1^E = 1 + b_1, \quad 
  b_2^E = 2 b_2 + \frac{8}{21} b_1, \quad b_s^E = b_s - \frac{2}{7} b_1
  \quad .
\label{eq:bias}
\end{equation}
The linear Eulerian bias $b_1^E$, in particular, is commonly quoted in literature as ``the bias'' of a field (e.g.~$\delta_{\rm tracer} = b_1^E\,\delta + \cdots$).  We see that $b_1^E\approx 1$ implies $b_1\approx 0$\footnote{Note that $b_2^E\approx0$ does not imply that $b_2\approx0$, or vice versa, due to the $b_1$ term in the transformation. This implies that $\delta_e=b_{1,e}^E\delta_m$ leads to $r_{em}=1$, but not $b_{n>1,e}=0$.  Similarly $b_2=0\not\Rightarrow b_2^E=0$ if $b_1\ne 0$. }.

Power spectrum suppression is often quoted, measured and modeled in comparison to dark matter only simulations.  The degree to which this is necessary in fitting data is uncertain, but we provide a conversion here (see refs~\cite{Leung25,Hadzhiyska26} for more details).  Define $S(k)=P_{\rm full}(k)/P_{\rm dmo}(k)$, where $P_{\rm full}$ is the matter power spectrum including `all' of the physics and $P_{\rm dmo}$ is the same thing in a dark-matter-only simulation.  Ignoring neutrinos for now, decompose $\delta_m=(1-f_b)\delta_c+f_b\delta_b$ where $f_b\simeq 0.16$ is the baryon fraction.
In the limit that the impact of feedback on the CDM is small we can replace $P_{cc}$ with $P_{\rm dmo}$ and on scales where baryons remain well correlated with matter ($r_{bc}\simeq 1$) \cite{Hadzhiyska26,Ebina26d}
\begin{equation}
    P_{mm} = (1-f_b)^2 P_{\rm dmo} + 2f_b P_{bm} - f_b^2 P_{bb}
    \quad\Rightarrow\quad
    S(k) \simeq (1-f_b)^2 + 2f_b\frac{P_{bm}}{P_{\rm dmo}} - f_b^2 \frac{P_{bc}^2}{P_{\rm dmo} P_{cc}}
\end{equation}
We shall assume that ionized electrons trace the baryons up to a small correction due to stars, $f_\star$, so that $P_{em}=(1-f_\star)P_{bm}$.  We can further replace $P_{bc}$ by $P_{bm}$ and $P_{cc}$ by $P_{\rm dmo}$ with little error so 
\begin{equation}
    S(k) \simeq (1-f_b)^2 + 2\tilde{f}_b\frac{P_{em}}{P_{\rm dmo}} - \tilde{f}_b^2 \left( \frac{P_{em}}{P_{\rm dmo}} \right)^2
    \label{eq:Sk_Pem}
\end{equation}
where $\tilde{f}_b=f_b/(1-f_\star)$\footnote{If we had used $r_{bm}=1$ instead of $r_{bc}=1$ (which are equivalent statements) the third term in Eq.~\ref{eq:Sk_Pem} would become $-\tilde{f}_b^2 P_{em}^2/P_{\rm dmo}P_{mm}$ with a different shape at small scales, although this is suppressed by $\tilde{f}_b^2\approx0.025$.}. Note that recent work has identified the stellar fraction as a key uncertainty in interpreting kSZ measurement at one-halo scales \cite{Liu26}, which may impact the approximation above.

\section{Forecasting formalism}
\label{sec:forecasting}

Given the models above, we can estimate the degree to which we can constrain the suppression of the matter power spectrum from baryon feedback, given observations of FRBs and galaxies, using the Fisher formalism \cite{Fisher,frieden98,Coe09}. 

\subsection{Fisher formalism}

Suppose we measure some data (in this case the galaxy auto-spectra and the cross-spectrum with the FRB dispersion measure) each with some accuracy. Let $C$ be the covariance matrix for the stacked data vector of these observations.  Further assume that we have a model, $m_i$, capable of describing these data given a set of parameters, $p_a$.  The Fisher matrix is then \cite{Fisher,Tegmark97b}
\begin{equation}
      F_{ab} = \frac{dm_i}{dp_a} \, C_{ij}^{-1} \, \frac{dm_j}{dp_b}
      \quad .
\end{equation}
Inverting $F_{ab}$ provides an estimate for the parameter covariance matrix, $C_{ab}$. Given a prior on any of these parameters we add the inverse of the prior covariance to $F_{ab}$ before inverting the sum. For small deviations from a fiducial model, linearizing the response of $P_{mm}(k;p)$ gives
\begin{equation}
  \mathrm{var}\left[ P_{mm}(k_i) \right] = \frac{d\tilde{m}_i}{dp_a}\,  C_{ab}\, \frac{d\tilde{m}_i}{dp_b}
\end{equation}
where $\tilde{m}_i$ refers to the model prediction of $P_{mm}(k_i)$, given the parameters.
This allows us to use the model, $m$, to translate our observational uncertainty in $gg$ and $ge$ into an implied uncertainty on $mm$.

\subsection{Input data}

\begin{table}
    \centering
    \caption{Fiducial galaxy and electron bias parameters used for Fisher forecasts. The second-order galaxy bias is set by Sheth-Tormen peak-background split and the counterterm and Fingers-of-God is set by fits to halo occupation simulations \cite{Sailer21}. 
    All biases are Lagrangian, each of which can be converted to Eulerian by the conversion shown in Eq.~\ref{eq:bias}. $b_\nabla$ is in $h^{-2}{\rm Mpc}^2$,
    $N_{gg}$ is in $h^{-3}{\rm Mpc}^3$. Higher-order electron biases, as well as the cross shot noise $N_{eg}$, are set to zero in agreement with previous simulation fits \cite{Ebina26d,Sullivan26}. All other parameters not shown here are set to zero by default. }
    \label{tab:params}
    \begin{tabular}{lcccccccc}
    \hline
    $z$ & $b_{1,e} = b_{1,e}^E - 1$ & $b_{\nabla,e}$ & $R_e$ & $b_{1,g} = b_{1,g}^E - 1$ & $b_{2,g}$ & $b_{\nabla,g}$ & $N_{gg}$ & $N_{2,gg}$ \\
    \hline
    $0.2$ & $-0.037$ & $0.566$ & $0.519$ & $0.490$ & $-0.525$ & $2.482$ & $1000$ & $-1171$ \\
    $0.5$ & $-0.041$ & $0.530$ & $0.455$ & $0.885$ & $-0.228$ & $3.646$ & $2529$ & $-3253$ \\
    \hline
    \end{tabular}
\end{table}
        
Most forecasts will utilize galaxy auto-correlation and galaxy-dispersion cross-correlation, at two redshift bins ($z=0.2$ and 0.5) with bin widths $\Delta z=0.2$ ($\pm0.1$) and 0.3  ($\pm 0.15$), respectively. This is motivated by the signal-to-noise anticipated from next-generation surveys (see e.g.~Fig.~2 of ref.~\cite{Ebina26d}), as well as the redshift ranges probed by weak lensing surveys. 
The lower redshift bin probes a smaller redshift range in order to limit the physical bin depth $\Delta\chi$ and to limit the low-$\chi$ contribution which is substantial for $D$, as discussed below. 
In addition, we will run forecasts assuming an idealized shear survey. First we describe the FRB and galaxy surveys central to this work. 

For the FRB data, we assume scenarios similar to next-generation experiments \cite{Hallinan19,CHORD,Connor26}, which will be significantly more powerful than the datasets available now.  We assume an FRB footprint similar to DSA, covering all of the Dark Energy Spectroscopic Instrument (DESI) \cite{DESI,DESI-DR1} footprint, corresponding to $f_{\rm sky}=0.41$ of common area.  Since the properties of these as-yet-to-be-discovered FRBs are highly uncertain, we shall take round numbers and explore how the constraints change under variations of our assumptions.  For our fiducial model we assume a net FRB density of $10\,\deg^{-2}$, which is then downsampled to FRBs that are behind each galaxy sample using the FRB redshift distribution in Fig.~\ref{fig:FRBz}.  
For this distribution $P(>z)=\exp[-z/z_0]\left(1+z/z_0+\tfrac{1}{2}[z/z_0]^2\right)$ and we have 81\% and 37\% of our full FRB population above $z=0.3$ and 0.65.  Thus we take $\bar{n}_{\rm eff}=8.1\,\mathrm{deg}^{-2}$ and $3.7\,\mathrm{deg}^{-2}$ for the $z=0.2$ and 0.5 $z$-bins, respectively.

For the galaxy samples, we will utilize those anticipated to be collected by DESI on timescales similar to DSA and CHORD. Specifically, we will use the Bright Galaxy Survey (BGS; \cite{DESI24-II}) at $z=0.2$ and the Luminous Galaxy Extension (LGE) from DESI Run 1b (expected to complete in 2028; \cite{DESI-Run1b}) at $z=0.5$.

As described in \S\ref{sec:theory}, the electron biases probed by FRB DM are constrained by any deviation from $r_{em}=1$. For the fiducial biases of electrons we assume all of the higher-order Lagrangian biases are zero and $b_{1,e}^L$ is only a few percent. For feedback and small-scale biases, we adopt values based on fits to the FLAMINGO simulation \cite{Schaye23,Kugel23,Helly26} fiducial feedback model (\texttt{L1\_m9}) using HEFT, up to $k=2\,h\,\mathrm{Mpc}^{-1}$, where we are able to fit to significantly smaller scales than standard HEFT by excluding higher-order bias contributions. We refer the reader to  Table~\ref{tab:params} for details and Fig.~\ref{fig:Pee_damping} for the HEFT fit. 

We shall take the linear biases of the BGS and LGE samples from those of existing DESI samples.  For the BGS we use $b_1^E(z)=1.34/D(z)$ and for the LGE, we assume they have the same bias as the LRG sample: $b_1^E(z)=1.45/D(z)$ \cite{DESI24-II}.  Recall $b_1^E=b_1(z)+1$.
We predict the second-order bias using the Sheth-Tormen peak-background split \cite{Sheth99} and model the galaxy counterterms with fits to halo occupation simulations as in refs.~\cite{Sailer21,Ebina24a}. For the fingers-of-God (FoG) term, we adopt a fiducial value of $N_2=-\sigma_v^2/\bar{n}$, where $\sigma_v=100\,\mathrm{km}/\mathrm{s}$ is a typical galaxy velocity dispersion \cite{Sailer21,Ebina24a}. All other parameters are marginalized around zero. Table~\ref{tab:params} lists the fiducial bias values.

In the course of this work, we also consider other spectroscopic catalogs, such as the Extreme Emission Line Galaxies (XLGs) from DESI Run 2 (start 2029; \cite{Schlegel22}), which should be available on a timescale similar to that of next-generation FRB surveys. Although XLGs have a significantly higher number density than BGS over comparable redshift ranges and are signal dominated to similar $k$, they yield moderately weaker constraints because of their smaller linear bias. As discussed in \S\ref{sec:results}, we attribute much of this difference to the lower value of $b_{2,g}$, and we expect a similar trend to hold for other low-bias tracers.

Using these galaxy and electron samples, we will use the redshift-space galaxy auto-spectrum $P_{\ell}^{gg}$, angular galaxy auto-spectrum $C_\ell^{gg}$, and galaxy-dispersion cross-spectrum $C_\ell^{Dg}$ as the forecast observables. For $P_{\ell}^{gg}$, we adopt an EFT-like scale cut of $k_\mathrm{max}=0.2\,h\,\mathrm{Mpc}^{-1}$, while for angular spectra we adopt $\ell_\mathrm{max}$ equivalent to $k_\mathrm{max}=0.6\,h\,\mathrm{Mpc}^{-1}$, motivated by HEFT modeling limits. Furthermore, in order to neglect covariance between $P_{\ell}^{gg}$ and $C_\ell^{gg}$, we remove modes with  $k_{\parallel}<10^{-3}\,h\,\mathrm{Mpc}^{-1}$, excluding regions of $\mu\approx0$. This cut excludes a small region of 3D clustering and has a marginal effect. 

\subsection{Covariance}
\label{sec:covariance}

For the error on angular clustering we employ the Gaussian (or disconnected) approximation
\begin{align}
    \mathrm{Cov}\left[C_\ell^{gg},C_{\ell'}^{gg}\right] &=
    \frac{2\delta_{\ell\ell'}}{(2\ell+1)f_{\rm sky}\Delta\ell}
    (C_\ell^{gg}+N_\ell^{gg})^2 \\
    \mathrm{Cov}\left[C_\ell^{Dg},C_{\ell'}^{Dg}\right] &=
    \frac{\delta_{\ell\ell'}}{(2\ell+1)f_{\rm sky}\Delta\ell}
    \left[(C_\ell^{DD}+N_\ell^{DD})(C_\ell^{gg}+N_\ell^{gg}) + (C_\ell^{Dg})^2\right]
\label{eq:CovCDg}
\end{align}
which is expected to dominate on large scales.  Appendix \ref{sec:zweight} discusses the integration over the FRB redshift distribution relevant for the $Dg$ cross spectrum covariance.  We shall use the noise-dominated limit throughout.

We equivalently take the Gaussian approximation for the redshift-space power spectrum $P_\ell^{gg}$
\begin{equation}
\mathrm{Cov}\left[P^{gg}(\mathbf{k}),P^{gg}(\mathbf{k}')\right] =
2\delta_{\mathbf{k},\mathbf{k}'}
\left[P^{gg}(\mathbf{k})+\frac{1}{\bar n_g}\right]^2
\label{eq:CovPggRSD}
\end{equation}
where, as mentioned above, the cross-covariance between 3D and angular clustering is neglected by removing modes at $\mu\approx 0$. 

\subsection{Scale sensitivity}
\label{sec:frb_ksz_comparison}

\begin{figure}
    \centering
    \includegraphics[width=1.0\linewidth]{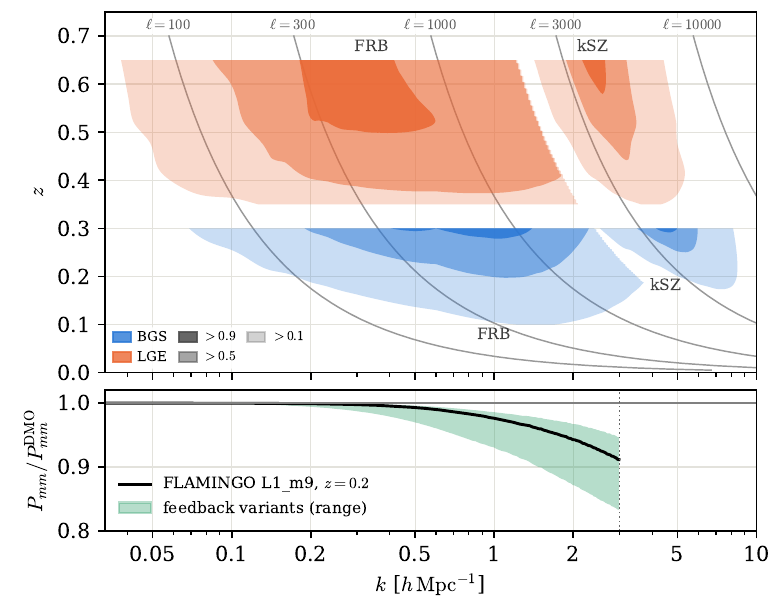}
    \caption{Top: Fisher forecast constraints on a modulation of $P_{ge}$ at a certain scale and redshift.  We plot the inverse variance of a fractional perturbation to $P_{ge}$, as described in the text.  In both cases the redshift range is selected by the limits of the galaxy samples (we show BGS and LGE). 
    The FRB dispersion-galaxy cross-spectrum constrains significantly larger scales (lower $k$) than those probed by the kSZ effect as well as a wider logarithmic range in $k$.
    The contour shades show 90\%, 50\%, and 10\% of the peak, with each contour normalized separately. The small-scale limit of the FRB contour is defined by $\ell_{\rm max}=2000$ adopted in this work and can in principle be extended, with limited modeling control. 
    Bottom: The power spectrum ratio, $S(k)=P_{\rm full}(k)/P_{\rm dmo}(k)$, for our fiducial FLAMINGO simulation at $z=0.2$ (black line).  The shaded band shows the range of variation based on the feedback variant modeled.  
    The vertical dotted line shows the approximate resolution limit of the FLAMINGO simulation.}
    \label{fig:scale-redshift}
\end{figure}

Using the Fisher formalism and Gaussian covariance matrices described above, one can determine the scale and redshift sensitivity of FRB-galaxy and kSZ measurements. 
Let us perturb $P_{ge}$ in a small region around $(k,z)$ with fractional amplitude $\Delta$, i.e.\ $P_{ge}(k,z)=P_{\rm fid}\left[ 1 + \Delta\ \mathcal{B} \right]$ where $\mathcal{B}(k,z)$ is a top-hat function over a small region centered on $(k,z)$.  Assuming $\Delta$ is small and this is the only change to $P_{ge}$, the degree to which we can constrain $\Delta$ is simply given by
\begin{align}
  \sigma_{\Delta}^{-2} &= \frac{\partial m}{\partial\Delta} C^{-1} \frac{\partial m}{\partial\Delta}
\end{align}
where $m$ is our theoretical model for the data (e.g.\ $C_\ell^{Dg}$) and $C$ is the covariance. Using Eq.~\ref{eq:CovCDg} for the covariance of $C_\ell^{Dg}$ and making the identical approximation for the kSZ signal $C_\ell^{\pi T}$ \cite{Smith18,Qu26}, one obtains the sensitivity
\begin{align}
  \frac{\sigma_{\Delta}^{-2}}{(\delta z)^2 \delta \ln k} &=\frac{k}{\chi^3(z)H^2(z)} N_{\ell=k\chi} W_a^2(\chi) W_b^2(\chi) P_{\rm fid}^2(k,z)
\end{align}
after normalization to remove the dependence on the width of the top hat $\mathcal{B}$, with $W_a$ and $W_b$ the lensing kernel of the involved tracers ($D$ and $g$, or $\pi$ and $T$) and $N_\ell \,\Delta \ell=C^{-1}$. This is shown in Fig.~\ref{fig:scale-redshift}.  In both cases the redshift range is determined by the galaxy sample in the cross-correlation.  The FRB cross-correlation probes both significantly larger scales than the corresponding kSZ signal and also a wider range of scales (in a logarithmic sense).

\section{Results}
\label{sec:results}

Here we present forecast results using the formalism presented above. We will choose to interpret results as constraints on $S(k)=P_{mm}/P_{\rm dmo}$, a common metric to discuss control over feedback (see \S\ref{sec:theory}). As a reference point we note that forecast work on Stage~IV surveys often find $\mathcal{O}(1\%)$ control of $S(k)$ desirable at `feedback scales', if not better \cite{vanDaalen11,Chisari19,Amon22,Preston25,Truttero25,DeRose26c}. Throughout, we take $\ell_{\rm max}=2000$ as the small-scale limit of $C_\ell$ data vectors. This is beyond the HEFT modeling scale cut $k_{\rm max}$ and thus only affects the FRB-lensing forecasts. The qualitative conclusions there are insensitive to the precise choice of $\ell_{\rm max}$. The FRB-lensing forecast is summarized in Fig.~\ref{fig:CDk_forecast} and FRB-galaxy forecasts are summarized in Fig.~\ref{fig:forecast_summary} and Table~\ref{tab:forecast_summary}.

\subsection{FRB-lensing cross-correlation}

As a touchstone for which to compare the results in the following sections, let us first consider the case where we cross-correlate the FRB DM with a measure of the projected matter distribution, such as might arise from cosmic shear (see refs \cite{Leung25,Feng26,Sharma26,Sharma26b,Wayland26,Reischke26} for earlier work in this area).  We shall assume that tomographic information on the shear is available so that it is possible, at least in principle, to select the contributions from a specific redshift slice.  Our goal here is not to forecast a particular survey, or the influence of systematic errors on inference, rather it is to establish a ``base case'' to which we can compare in the next sections.

To this end we assume an ideal shear sample, with one source redshift $z_s$, with a source catalog density of $27\,\mathrm{arcmin}^{-2}$ equivalent to the total LSST Y10 sample \cite{LSST}, and an FRB distribution of known $dN/dz$ according to Fig.~\ref{fig:FRBz}. We choose $z_s=0.6$ and 1.0 in order to retain FRBs with $z<z_s$ for cross correlation, while allowing us to later neglect FRB-source overlap by neglecting FRBs with $z>z_s$. 
While idealized, the source catalog redshift is within the expected range of catalogs from LSST and Euclid.

The forecast formalism is nearly identical to that of FRBs, but with the shear signal probing the integrated matter density
\begin{equation}
    \kappa_i = \int_0^{\chi_s} d\chi \, W_\kappa(\chi) \, \delta_m(\chi \hat{\mathbf{n}}),
\end{equation}
with the redshift kernel $W_\kappa$ peaking halfway between the observer and source. 
This can be extended to a general shear catalog with known distribution $n_i(\chi)$, 
\begin{equation}
    W_i(\chi) = \tfrac{3}{2}\Omega_m H_0^2 \frac{\chi}{a}\int_\chi^\infty d\chi'\, n_i(\chi')\frac{\chi'-\chi}{\chi'}
\end{equation}
which widens the kernel, diluting the SNR and de-localizing the measurement. 

\begin{figure}
    \centering
    \includegraphics[width=1.0\linewidth]{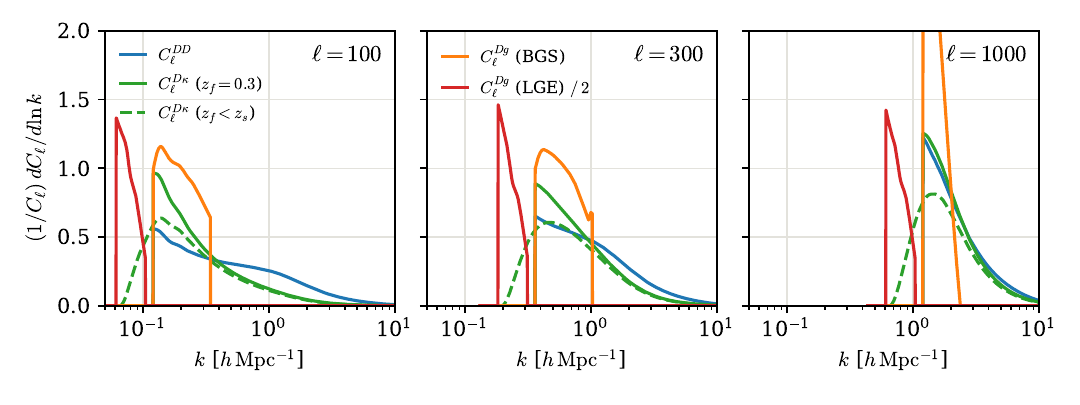}
    \caption{The information content by wavevector of $C_\ell^{DD}$ for $z_f=0.3$ (blue), $C_\ell^{D\kappa}$ for $z_s=0.6$ (green), and $C_\ell^{Dg}$ for BGS (orange) and LGE (red) samples. The solid and dashed green lines correspond to assuming FRBs at $z_f=0.3$, or using all FRBs with $z_f<z_s$ according to the FRB distribution in Fig.~\ref{fig:FRBz}. The latter case corresponds to the $C_\ell^{D\kappa}$ forecast in the main text.
    $C_\ell^{DD}$ extends to high-$k$ the most given the large contribution at low $\chi$. $C_\ell^{D\kappa}$ similarly extends to high-$k$, but with significantly smaller contributions as $W_\kappa(\chi\to0)\propto\chi$ and with low $k$ modes contributing when the FRB distribution is of finite width. 
    The DM-galaxy cross correlations, $C_\ell^{Dg}$, are `localized' by the galaxy redshift bins, giving rise to discontinuous curves and a tighter mapping between $\ell$ and $k$.}
    \label{fig:Cl_dlnk}
\end{figure}

There are multiple complications to this measurement related to the source catalog, including its intrinsic alignment and uncertainty in the redshift distribution, and its finite redshift distribution width de-localizing cross-correlation with FRBs. 
By assuming a single source redshift $z_s$ and limiting the FRB catalog to those in front of the source catalog, we are able to avoid or mitigate the concerns above. 

Even in this scenario, modeling and interpretation of $C_\ell^{D\kappa}$ is non-trivial. A practical complication is that in this case, there is an overlap between the lensing kernel $W_\kappa$ and the redshift distribution of FRBs. This requires near-perfect knowledge over the FRB redshifts, a significantly stronger assumption than knowing that FRBs are closer or further than some redshift, and also requires control over the host dispersion measure contribution. 
Another issue is structural, due to the shape of the composite kernel $W_DW_\kappa$. As the DM kernel does not approach zero as $\chi\to 0$, the overall factor of $\chi^{-2}$ in the spectra leads to significant contributions from low redshift, or equivalently high $k$. 
Fig.~\ref{fig:Cl_dlnk} shows that at even moderate $\ell$ there is significant signal from high $k$ ($k\gtrsim1\,h\,\mathrm{Mpc}^{-1}$), where the impact of feedback is poorly understood. 
In addition, the lack of redshift localization in the composite kernel $W_D W_\kappa$ means the signal comes from a wide range of redshifts (with emphasis on low redshifts), such that any redshift evolution of feedback or electron bias would need to be included, making it difficult to constrain feedback at redshifts relevant to shear surveys. 

Thus, while in principle the matter field sensitivity of lensing allows us to directly probe $P_{me}$ through cross-correlation with FRBs (which can be converted to $S(k)$ using the expression in Eq.~\ref{eq:Sk_Pem} with the assumptions $r_{bc}\simeq 1$, $P_{cc}\approx P_{\rm dmo}$ and a small correction due to baryons in stars \cite{Leung25,Hadzhiyska26}), in practice the issues above, and in particular the lack of scale and redshift localization, render this difficult.
Instead, we choose to parametrically forecast our constraints on $S(k)$, using the bias and feedback models from \S\ref{sec:theory}. We only include linear electron bias $b_{1,e}$ and feedback parameters $b_{\nabla,e}$ and $R_e$ for an optimistic forecast. 
We consider the following variations, to account for and assess the difficulties of modeling $C_\ell^{D\kappa}$. 
As an ideal case, we assume that none of the parameters evolve with redshift, and that our feedback model is accurate to very small scales (which can enter the signal, as shown in Fig.~\ref{fig:Cl_dlnk}). 
Then, to account for the scale limits of our feedback model, we introduce freedom in the high-$k$ spectrum. Specifically we marginalize over templates $T(k)\subset P_{em}(k)$ such that $T(k<k_c)=0$, where $k_c$ is a cut-off scale below which we think our model is adequate.  We optimistically set $k_c=2\,h\,\mathrm{Mpc}^{-1}$, loosely motivated by hydrodynamical simulations (see e.g.~Fig.~\ref{fig:Pee_damping}). 
Furthermore, we consider a linear redshift evolution of electron bias and feedback parameters, to account for the redshift localization of the signal. 

\begin{figure}
    \centering
    \includegraphics[width=1.0\linewidth]{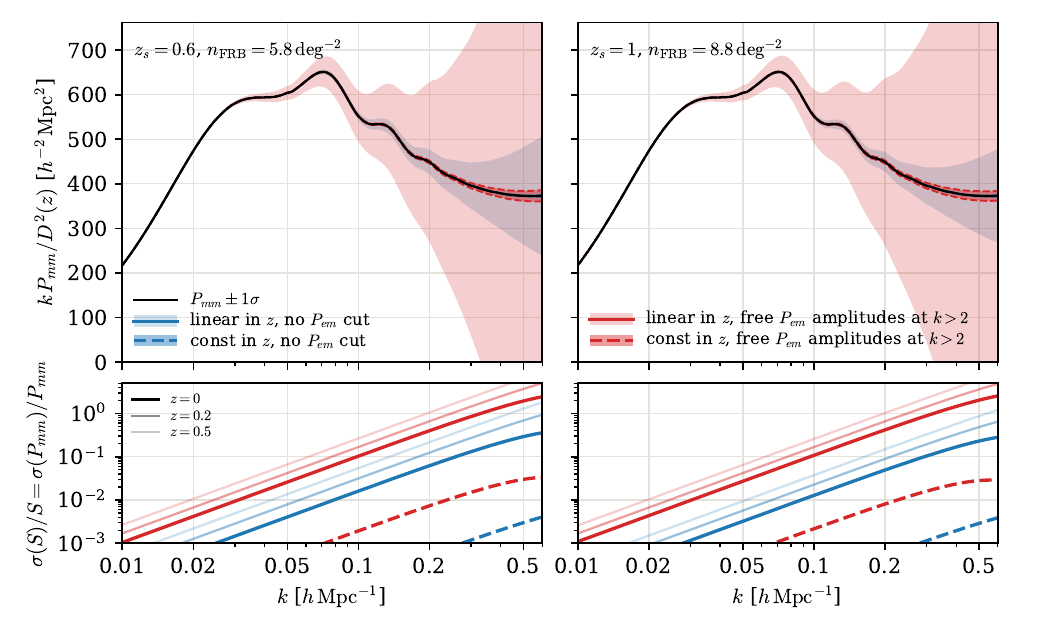}
    \caption{Forecasts based on FRB dispersion-shear cross correlation, assuming an idealized shear source sample at a single redshift $z_s$ and density equivalent to Y10 LSST. The most idealistic forecast obtains sub-percent measurements of $S$, while the lack of modeling at very small scales and redshift evolution of electron parameters (bias and feedback) significantly degrade the forecast. See main text for other complications not forecast here. 
    }
    \label{fig:CDk_forecast}
\end{figure}

The forecasts, including all variants described above, are shown in Fig.~\ref{fig:CDk_forecast}. 
In the most optimistic scenario, considering constant parameters and information across all $k$, one obtains 0.4\% (0.4\%) constraints on $S(k=0.6\,h\,\mathrm{Mpc}^{-1})$ at $z_s=0.6$ (1.0), reaching the $\mathcal{O}(1\%)$ constraints required for shear surveys. This degrades by a factor of $\sim8$ when excluding information above $k_c=2\,h\,\mathrm{Mpc}^{-1}$, to 3.3\% (2.9\%) at $z_s=0.6$ (1.0). 
When allowing for redshift evolution of bias and feedback parameters, the constraints significantly degrade. With a linear evolution and information from $P_{em}$ across all $k$, one constrains $S(k=0.6\,h\,\mathrm{Mpc}^{-1})$ at $z=0$ at 36\% and 28\% with $z_s=0.6$ and 1.0, nearly two orders of magnitude worse than when considering redshift invariant parameters. This further degrades when constraining $S(k)$ at $z=0.2$ and 0.5, to 93\% (65\%) and 180\% (120\%) at $z_s=0.6$ (1.0), significantly above the required precision for shear surveys. 
This further worsens by a factor of five or more when excluding information past $k_c$ with linearly evolving parameters\footnote{In principle, it would be consistent to consider the evolution of high-$k$ templates with redshift as well, which will affect the constraints even more.}.
While tomographic FRB bins may improve constraints on the linear parameter evolution, it is unlikely that these will yield $\mathcal{O}(1\%)$ measurements. 
Note again that we have made multiple idealizations in this forecast as mentioned above, and we neglect higher-order modeling of electrons that can further degrade constraints, as we describe below.

\subsection{FRB-galaxies with linear galaxy bias}

A cross-correlation of FRBs and galaxies is another approach to constrain electron clustering, which mitigates many of the caveats of FRB-shear clustering. 
The simplest case is to assume that both electrons and galaxies faithfully trace the matter field (and hence each other) up to feedback scales.  This would be the case if both FRBs and galaxies were linearly biased tracers of the matter field.  While this is close to true for FRBs, it is known not to hold for galaxies.  However the case is instructive nonetheless.

If $\delta_g=b_g\delta_m$ and $\delta_{e}=b_e\delta_m$ then we can measure $P_{ee}$ via the combination $P_{ge}^2/P_{gg}$, without needing to measure $P_{ee}$ directly.  In reality one would be unlikely to `divide by $P_{gg}$' due to the poor noise properties of this quantity, however the same effect could be accomplished by a fit of a parameterized model to the joint data set: $P_{ge}$ and $P_{gg}$.  This has the advantage of leveraging the low-noise properties of the galaxy field, which will likely remain much better measured than the FRBs for some time, while simultaneously allowing isolation of particular redshift intervals.
Using a combination of the redshift-space $P_{gg}$, $C_\ell^{gg}$ and $C_\ell^{Dg}$, with a 2\% prior on $f_e$, one obtains 1.6\% and 1.8\% constraints on $S(k)$ at $k=0.6 \,h\,\mathrm{Mpc}^{-1}$, at $z=0.2$ and 0.5 respectively (blue in Fig.~\ref{fig:forecast_summary}). 
This optimistic scenario roughly matches the $\mathcal{O}(1\%)$ constraints on $S(k)$ often assumed for a data-limited analysis of Stage~IV lensing surveys \cite{vanDaalen11,Chisari19,Amon22,Preston25,Truttero25,DeRose26c}.

\subsection{Beyond linear galaxy bias}

\begin{figure}
    \centering
    \includegraphics[width=1.\linewidth]{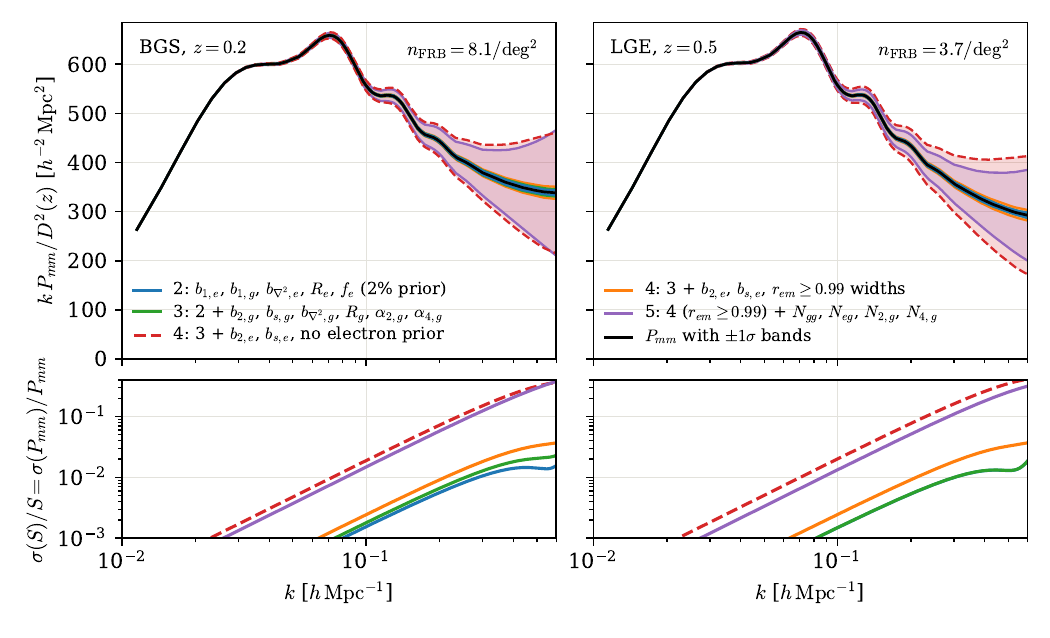}
    \caption{Forecasts for $P_{mm}$ and error on $S(k)$ from a galaxy-electron joint fit using $P_{\ell,gg}$, $C_\ell^{gg}$, and $C_\ell^{Dg}$ described in \S\ref{sec:forecasting} and \S\ref{sec:results}. The top panels show the $1\sigma$ error bands around the fiducial $P_{mm}$, while the bottom panels show the fractional error on $S(k)$. The scenario numbers in the legend correspond to the respective subsections of \S\ref{sec:results}.}
    \label{fig:forecast_summary}
\end{figure}

Of course, galaxy bias is not scale-independent \cite{Barnes1985,White1987} and this causes the galaxy, FRB and matter fields to decorrelate (Eq.~\ref{eq:rem_taylor}). We can handle this in our forecasts by introducing $b_{n>1}$ for the galaxy field, and allowing the data to determine them along with the feedback parameters.
This has varying impacts for the forecasts at each redshift bin. At $z=0.2$, the constraining power is degraded, to 2.3\% at $k=0.6\,h\,\mathrm{Mpc}^{-1}$, whereas that at $z=0.5$ is affected marginally, yielding 1.8\% constraints (green on Fig.~\ref{fig:forecast_summary}). 

The degradation is a strong function of the fiducial $b_{2,g}$ value. Choosing $b_{2,g}\approx0$ at $z=0.2$, rather than $b_{2,g}\approx-0.5$ as predicted by Sheth-Tormen, gives results similar to assuming linear galaxy bias, with a 1.7\% constraint on $S(k=0.6\,h\,\mathrm{Mpc}^{-1})$. This is reflective of the increased degeneracy between $b_{2,g}$ and $b_{\nabla,g}$ for low $b_{2,g}$, which significantly degrades constraints on feedback. 
While the true value of $b_{2,g}$ is outside of our control, it is known that higher-order statistics, particularly the bispectrum, marked power spectrum, or the (simple) skew spectrum, significantly tighten constraints on $b_{2,g}$ \cite{Ivanov22a,Chudaykin25a,Ebina24,Ebina26b}. A threefold increase in the constraining power of $b_{2,g}$, for example, would constrain $S(k=0.6\,h\,\mathrm{Mpc}^{-1})$ to 1.7\% at $z=0.2$ with our fiducial $b_{2,g}$, matching the linear galaxy bias scenario. 

\subsection{Beyond linear FRB bias}

At present we have no empirical evidence that the FRB bias is scale-independent, only arguments from theory or measurements from simulations \cite{Ebina26d,Sullivan26}.  Thus, here we consider to what extent non-linear bias for the FRBs would be important.  A useful metric of the strength of the prior we place here is the (de-)correlation of electrons and matter, quantified through $r_{em}$.  This can be converted to an upper bound on how large the Lagrangian biases can be using Eq.~\ref{eq:rem_taylor}. 

Having considered $b_{n>1,e}=0$ above, here we consider only two additional cases: one using a 1$\sigma$ prior on higher-order electron biases set by $r_{em}\ge0.99$ at $k<0.3\,h\,\mathrm{Mpc}^{-1}$, and one allowing complete freedom for the electron biases. Even in the former scenario, where we assume up to 1\% de-correlation at quasi-linear scales\footnote{While this is a powerful assumption, this is less stringent than measurements from some simulations including FLAMINGO, which find $r_{em}\gtrsim0.999$ at $k<0.3\,h\,\mathrm{Mpc}^{-1}$ \cite{Ebina26d}. Here we consider $r_{em}\ge0.99$ given its deviation from the linear electron bias case earlier and difficulty simulating baryons, even at quasi-linear scales \cite{Sharma25}.}, we find 3.7\% constraints on $S(k=0.6\,h\,\mathrm{Mpc}^{-1})$ at both $z=0.2$ and 0.5, significantly worse than assuming linear electron bias (orange in Fig.~\ref{fig:forecast_summary})\footnote{In this case, the $z=0.2$ constraint improves negligibly with a higher value of $b_{2,g}$, demonstrating that the control on electron biases dominates over that of $b_{2,g}$.}. 
In the latter scenario, the constraints worsen by an additional order of magnitude, with 36\% and 41\% constraints on $S(k=0.6\,h\,\mathrm{Mpc}^{-1})$ at $z=0.2$ and 0.5 (red dashed in Fig.~\ref{fig:forecast_summary}).

\subsection{Stochasticity of electrons and galaxies}

Finally, we include the stochasticity of the electron and galaxy fields.  Within EFT the `stochastic' component refers to the impact of small-scale perturbations that are uncorrelated with the large scale density field \cite{Ivanov22b}.  In the absence of conservation laws (like mass or momentum conservation) the lowest order term is white noise.  The galaxy stochasticity is dominated by shot noise and finger-of-god contributions. The electron-galaxy cross-stochasticity is also expected to appear in $C_\ell^{Dg}$ as a shot noise contribution, $N_{eg}\subset P_{eg}$.  A rough estimate for $N_{eg}$ would be $\sqrt{N_{ee}N_{gg}}$, with a prefactor depending upon the degree of overlap of the host halos for the populations.  Naively this would be small due to two factors.  First the lack of overlap in host halos (or the lack of host halos for electrons) would suppress any cross-stochasticity, e.g.~$N_{eg}\ll\sqrt{N_{ee}N_{gg}}$.  Second $N_{ee}$ could be very small if it is dominated by electron shot noise.  This is what we have considered thus far.  Joint electron-galaxy EFT fits to the FLAMINGO simulation found $N_{eg}\sim\sqrt{N_{ee}N_{gg}}$, with the relevant $N_{ee}$ being set by simulation resolution, while being consistent with zero to within $1.5\sigma$ \cite{Ebina26d}.  This indicates that even in the simulation-based landscape our knowledge of cross-stochasticity is limited.  Since stochasticity between tracers is often complex and an intuitive picture is insufficient for us to neglect even the possibility of cross-stochasticity, we provide forecasts including $N_{eg}$.  The forecasts demonstrate that the cross-stochasticity can significantly degrade constraints, showing 38\% and 31\% constraints on $S(k=0.6\,h\,\mathrm{Mpc}^{-1})$ at $z=0.2$ and 0.5 respectively, with the $r_{em}\ge0.99$ prior (purple in Fig.~\ref{fig:forecast_summary}).  Further investigation of this topic is clearly warranted.

\begin{table}
    \centering
    \begin{tabular}{l cccc}
    \hline
     & \multicolumn{2}{c}{BGS, $z=0.2$} & \multicolumn{2}{c}{LGE, $z=0.5$} \\
    case & $\sigma(b_{\nabla,e})$ & $\sigma(R_e)$ & $\sigma(b_{\nabla,e})$ & $\sigma(R_e)$ \\
    \hline
    2: $b_{1,e}, b_{1,g}, b_{\nabla,e}, R_e, f_e$ (2\% prior)                           & 2.03 & 10.7 & 1.96 & 13.5 \\
    3: 2 + $b_{2,g}, b_{s,g}, b_{\nabla,g}, R_g, \alpha_{2,g}, \alpha_{4,g}$             & 2.35 & 11.1 & 1.97 & 13.6 \\
    4: 3 + $b_{2,e}, b_{s,e}$, $r_{em} \geq 0.99$ widths                              & 3.18 & 12.4 & 3.20 & 15.9 \\
    4: 3 + $b_{2,e}, b_{s,e}$, no electron prior                                      & 24.4 & 57   & 26.0 & 69   \\
    5: 4 ($r_{em} \geq 0.99$) + $N_{gg}, N_{eg}, N_{2,g}, N_{4,g}$                    & 19.1 & 27   & 17.1 & 36   \\
    \hline
    \end{tabular}
    \caption{The error on feedback parameters from the forecast of \S\ref{sec:results} (fiducial values can be found in Table \ref{tab:params}). The case numbers correspond to the respective subsections of \S\ref{sec:results}. Note that the error on $R_e$ exceeds the expected $\mathcal{O}(1\,h^{-1}\,\mathrm{Mpc})$, especially in the latter two cases. Placing a $\mathcal{N}(0,3)$ prior on $R$, however, only noticeably changes the error on $S(k)$ in the latter two cases when the error is already $>10\%$ and hence has little influence on the qualitative conclusions of this work.}
    \label{tab:forecast_summary}
\end{table}

\section{Conclusion}
\label{sec:conclusion}

Fast radio burst (FRB) dispersion measures (DM) offer a window into the late-time distribution of baryons. With the rapidly expanding experiment capabilities this will allow us to probe baryonic feedback, one of the leading systematics for Stage-IV weak lensing surveys and a critical ingredient in galaxy formation. In this work, we employ hybrid effective field theory (HEFT) and the Fisher matrix formalism to investigate how `external' data from cross-correlations of FRBs with optical surveys can be used to place priors/constraints on the matter power spectrum suppression from baryonic physics.

By modeling the way in which galaxies trace the matter field using the bias expansion, tied to the fundamental symmetries of the problem rather than a particular model of galaxy formation, with parameters tightly constrained by galaxy clustering data we remove much of the galaxy modeling uncertainty.  This model works best on precisely those scales ($0.1<k<1\,h\,\mathrm{Mpc}^{-1}$) where feedback has the greatest impact for Stage-IV weak lensing surveys.  We find that DM can be used to provide tight constraints on baryonic feedback under restricted assumptions about the non-FRB tracer used in the cross-correlation.
We incrementally increase the modeling freedom in our forecasts to determine which ingredients lead to the largest increase in uncertainty.
The tight constraints survive even allowing significant freedom in the galaxy-dark matter connection through nonlinear and scale-dependent bias, since the galaxy clustering itself suffices to calibrate the relations.  The constraints degrade more significantly if we include freedom in scale-dependent electron bias and stochasticity and exceed the $\mathcal{O}( 1\%)$ goal for maximizing cosmological constraints from Stage IV weak lensing surveys.  Thus finding better constraints on the higher-order electron biases and electron-galaxy stochasticity are particularly crucial to obtaining tight feedback constraints from FRBs and galaxies. 
Currently, knowledge of these quantities is driven by the available hydrodynamic simulations.  Given the difficulty of baryon modeling and the required precision of feedback constraints a measurement-based bound on these quantities would be valuable.  Conversely the approach outlined in this paper should allow upcoming FRB-galaxy cross correlation measurements to be used as a constraint or calibrator in hydrodynamical simulations of galaxy formation that is less sensitive to the modeling of particular observational samples \cite{McCarty26}.

Looking ahead, the most natural approach, from the point of view of fitting data, would be to include the FRB-galaxy cross-correlation as one of the data vectors in a global fit.  The other galaxy measurements provide constraints on the galaxy-matter connection and the FRB-galaxy data provide constraints on baryonic feedback with all of the correlated uncertainties properly accounted for.  Given a model for the FRB-galaxy data vector that shares the same nuisance parameters and theoretical basis, this is straightforward.  This approach has the advantage of naturally extending to include cosmic shear data (for which the baryonic feedback constraint is desired) and galaxy-shear cross-correlations while simultaneously propagating all of the statistical uncertainties and degeneracies.  To the extent that galaxy-shear cross-correlations only improve the constraints on the bias parameters we expect this will not improve our inference of feedback, as the uncertainty over galaxy biases are subdominant to other factors.

While our focus here has been on the use of FRBs to place informative priors on baryon feedback for weak lensing analyses, we note that another route is through the kSZ effect \cite{kSZ}, stacked around galaxies measured in photometric or spectroscopic surveys \cite{kSZ,Smith18,Madhavacheril19,Schaan21,Amodeo21,Bigwood24,Hadzhiyska25,RiedGuachalla25}.
In this respect, the generalization of HEFT known as CHEFT \cite{Pellejero26} would allow a similar analysis with many of the same considerations applying.  By bringing galaxy clustering, weak lensing, FRBs and kSZ probes into a common model space, this enables joint analyses focused on both large and small scale feedback \cite{Madhavacheril19,Sharma26b,McCarty26}.  We defer this combination, along with an investigation of kSZ-related modeling uncertainties \cite{Ondaro26}, to future work.

\section*{Acknowledgements}

We thank Jamie Sullivan for helpful discussions and comments upon an early draft.
HE and MW were supported by the DOE. 
This research used resources of the National Energy Research Scientific Computing Center (NERSC), a Department of Energy User Facility. 
Claude Code (Opus $\le5.5$ and Fable $\le5.2$) was used for software implementation in this work. 
All writing and analysis choices were made by the authors, and the authors take full responsibility for the contents of this work.

\appendix

\section{Redshift weighting}
\label{sec:zweight}

Consider galaxies in a redshift slice $z_{g,{\rm min}}<z<z_{g,{\rm max}}$.  The FRBs in front ($z_{FRB}<z_{g,{\rm min}}$) of the slice do not contribute to the cross-correlation, $C_\ell^{Dg}$ but can contribute `noise'.  The FRBs at $z_{FRB}\gg z_{g,{\rm max}}$ have very little of their total DM coming from electrons in the slice, and thus also contribute a lot of uncorrelated `noise' to $C_\ell^{Dg}$.  To reduce this noise, let us consider breaking our FRB sample into FRBs in a number of redshift slices, labeled by $i$.  When constructing the DM map, we weight each slice by $w_i$.  Our goal is to minimize the cross-spectrum error, Eq.~\ref{eq:CovCDg}, holding $C_\ell^{Dg}$ fixed.  We can achieve this by setting $w_i=0$ for $z_{FRB}<z_{g,{\rm max}}+\varepsilon$, with the padding $\varepsilon$ designed to eliminate the source-clustering contribution to simplify the modeling.  The cross-spectrum remains unchanged if we set $\sum_i w_i=1$.  Obviously $C_\ell^{gg}$ is independent of $w_i$ so we need simply minimize
\begin{equation}
    \sum_{ij} w_i \left( C_{\ell,ij}^{DD} + N_{\ell,ij}^{DD} \right) w_j \qquad\mathrm{with}\quad \sum_i w_i=1
    \quad , \quad w_{i<z_{g,{\rm max}}} = 0
\end{equation}
with the obvious definition of $C_{\ell,ij}^{DD}$.  Since the contribution to $D$ is cumulative, the matrix $C_{\ell,ij}^{DD}$ has the property that $C_{\ell,i<j}^{DD}=C_{\ell,ii}^{DD}$ (and is symmetric).  Dropping $i<z_{g,{\rm max}}$ from the problem, the solution is
\begin{equation}
    w = \frac{C_{\rm tot}^{-1}\mathbf{1}}{\mathbf{1}C_{\rm tot}^{-1}\mathbf{1}}
    \quad , \quad C_{\rm tot}=C_\ell^{DD}+N_\ell^{DD}
\end{equation}
for the `active' slices (i.e.\ $z_{\rm FRB}>z_{g,{\rm max}}$) else zero.
In the limit that $C_\ell^{DD}\gg N_\ell^{DD}$ only the lowest redshift FRBs with $z_{FRB}>z_{g,{\rm max}}$ should be retained.  If $N_{\ell,ij}^{DD}\propto \delta_{ij}/\bar{n}_{{\rm eff},i}$ and $N_{\ell,ii}^{DD}\gg C_{\ell,ii}^{DD}$ the weights per redshift bin trace $\bar{n}_{\rm eff}(z)$, i.e.\ each FRB is weighted equally.  In either case the `$D$-auto' contribution to var$[C_\ell^{Dg}]$, i.e.\  $C_{\ell}^{DD} + N_{\ell}^{DD}$ in Eq.~\ref{eq:CovCDg}, is replaced by $1/\sum_{ij} C_{{\rm tot},ij}^{-1}$, with the matrix and the sum only over bins with $z_{FRB}>z_{g,{\rm max}}$.

This optimal weighting will only become relevant for future surveys with large numbers of FRBs and for low $\ell$.  Before that time the optimal strategy is to give equal weight to each FRB with $z_{FRB}>z_{g,{\rm max}}$.  This sets the weights equal to the fraction of such FRBs in redshift shell $i$.  The noise, $N_\ell^{DD}$, is $\sigma_D^2/\bar{n}_{\rm eff}$ with $\bar{n}_{\rm eff}$ the effective angular density of all such FRBs.  The $C_\ell^{DD}$ piece in Eq.~\ref{eq:CovCDg} becomes
\begin{align}
    C_\ell^{DD}
    &\to C_{\ell,11}^{DD} + \sum_{i=2}^N
    \left( C_{\ell,ii}^{DD}-C_{\ell,i-1\ i-1}^{DD} \right) f_{\ge i}^2 \\
    &\to C_{\ell}^{DD}(z_{g,{\rm max}}) + \int_{z_{g,{\rm max}}}^\infty dz_f\ f^2(>z_f) \frac{dC_\ell^{DD}}{dz_f} \\
    &\to \int_0^\infty dz_f\ f^2(>z_f) \frac{dC_\ell^{DD}}{dz_f}
\end{align}
with $f_{\ge i}=\sum_{k\ge i} f_k$ being the fraction of FRBs in bin $i$ or above and $f(>z_f)$ the fraction with redshift $>z_f$ with $C_\ell^{DD}(z_f)$ the spectrum for FRBs at redshift $z_f$, Eq.~\ref{eq:ClDD}.  The first expression can be used with finite sized redshift bins, as may be desirable observationally.  The second line takes the continuum limit.  In the last line we have taken $z_{g,{\rm max}}\to 0$, i.e.\ computed the dispersion for the full sample, so that agreement with previous work is clear \cite{Wayland26,Feng26,Wang26}.

\section{Small-scale electron density perturbations}
\label{sec:Pee}

\begin{figure}
    \centering
    \includegraphics[width=1.0\linewidth]{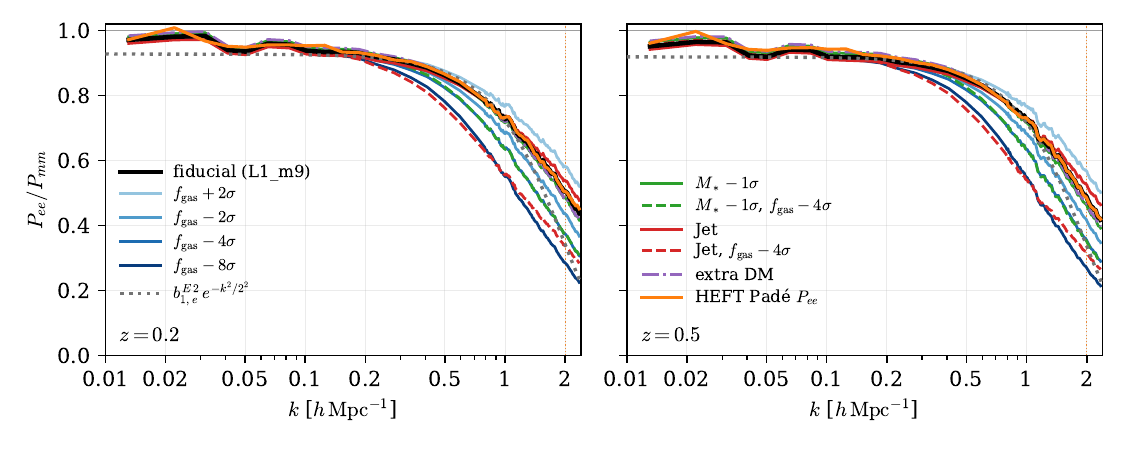}
    \caption{Ratio of electron and matter power spectra for various feedback models of the FLAMINGO simulation suite \cite{Schaye23,Kugel23,Helly26} at $z=0.2$ (left) and 0.5 (right). 
    The orange lines show the HEFT joint fit to electrons and CDM that is used as our fiducial model for $\delta_e$ (see Table \ref{tab:params} for the fit values).
    For the scales shown, the Gaussian suppression with a cutoff scale of $2\,h\,\mathrm{Mpc}^{-1}$ (dotted gray line) mimics the simulation curves well. }
    \label{fig:Pee_damping}
\end{figure}

The DM auto-spectrum is quite sensitive to small scale perturbations in the electron density field \cite{McQuinn14} due to the $\chi^{-2}$ weighting in the integrand of Eq.~\ref{eq:ClDD}.  This manifests as sensitivity to the high $k$ behavior of $P_{ee}$.  
When computing $C_\ell^{DD}$ we assume $P_{ee}$ is a linearly biased tracer, with the Pad\`e damping as described in the main text. 
We adopt the non-linear matter spectrum from HEFT to $k=3.5\,h\,\mathrm{Mpc}^{-1}$, which is further extrapolated using \texttt{HaloFit} \cite{Takahashi12}. 
Fig.~\ref{fig:Pee_damping} compares the HEFT linear bias fit, jointly fitting electrons and cold dark matter and conducted up to $k=2\,h\,\mathrm{Mpc}^{-1}$, to the $P_{ee}/P_{mm}$ suppression measured in variants of the FLAMINGO simulation suite \cite{Schaye23,Kugel23,Helly26} at $z=0.2$ and 0.5.  
The fit follows the fiducial \texttt{L1\_m9} model closely, despite fitting to $k=2\,h\,\mathrm{Mpc}^{-1}$, significantly past typical HEFT scale limits. This is enabled by deliberately constraining the non-linear electron biases to zero\footnote{We further constrain the electron shot noise $N_{ee}=0$, in order to suppress its disproportionately large effects at $k\gg1\,h\,\mathrm{Mpc}^{-1}$ and hence $C_\ell^{DD}$. Due to the high density of electrons this has negligible effect at our fit range. }.

\begin{figure}
    \centering
    \includegraphics[width=\linewidth]{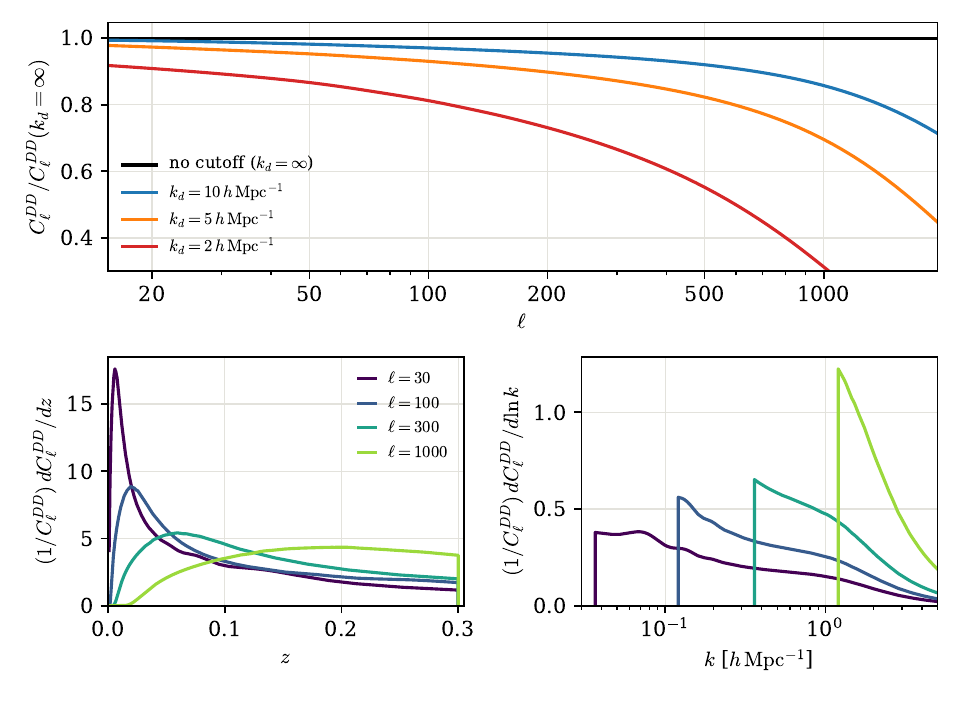}
    \caption{Top: Due to the kernel contribution extending to $\chi\to0$, FRB dispersion auto-spectrum $C_\ell^{DD}$ damps significantly as a result of a high-$k$ damping of $P_{ee}$. Even a cutoff scale $k_d=10\,h\,\mathrm{Mpc}^{-1}$, typically considered harmless for galaxy modeling, significantly damps the spectra at $\ell\gtrsim300$. Bottom: Information content of $C_\ell^{DD}$ by redshift (left) and scale (right). See also Fig.~\ref{fig:Cl_dlnk} for a comparison with $C_\ell^{D\kappa}$ and $C_\ell^{Dg}$. Note that for illustrative purposes, FRBs in this figure are sourced from a single redshift $z_F=0.3$, motivated by the peak of FRB distribution in Fig.~\ref{fig:FRBz}. }
    \label{fig:CDD_damping}
\end{figure}

Fig.~\ref{fig:CDD_damping} shows the impact of a Gaussian low-pass filter on $C_\ell^{DD}$, with a source FRB redshift of $z_F=0.3$ motivated by the peak of the FRB distribution in Fig.~\ref{fig:FRBz}.  
The suppression found in Fig.~\ref{fig:Pee_damping} is similar to $k_d=2\,h\,\mathrm{Mpc}^{-1}$, which leads to suppressions of tens of percent already at $\ell=100$ when compared to using $b_{1,e}^2P_{mm}$ with no filtering.  
Even a filter at $k_d=10\,h\,\mathrm{Mpc}^{-1}$, which would be considered `safe' for galaxy or cosmic shear spectra, shows $>10\%$ suppression at $\ell=10^3$. 
The lower panels of Fig.~\ref{fig:CDD_damping} show two different views of this phenomenon.  In the lower left we show $(1/C_\ell) dC_\ell/dz$ vs.\ $z$ to highlight just how much of the contribution in Eq.~\ref{eq:ClDD} comes from low $z$ (small $\chi$).  The lower right panel shows the complementary view of this, since small $\chi$ maps to large $k$ at fixed $\ell$.  Here we plot $(1/C_\ell)dC_\ell/d\ln k$ vs.\ $k$ for the same multipoles.  Note how much of the signal comes from high $k$ even for modest $\ell$.

Although $C_\ell^{DD}$ is highly sensitive to electron perturbations at high $k$ and low $z$, our headline results depend only weakly on this behavior. $C_\ell^{DD}$ enters our calculation only through the error on $C_\ell^{Dg}$ (Eq.~\ref{eq:CovCDg}) and they are dominated by uncertain FRB shot noise that exceeds the signal past $\ell\approx400$ even for $\bar{n}_{FRB}=10\,\deg^{-2}$ (lower right panel of Fig.~\ref{fig:FRBz}). As a result, a conservative small-scale suppression at $k_d=10\,h\,\mathrm{Mpc}^{-1}$ has only a subdominant effect for currently anticipated FRB samples. We therefore adopt $k_d=10\,h\,\mathrm{Mpc}^{-1}$ for the $P_{ee}$ modeling in the main text, to limit our dependence on the poorly known clustering at very small scales.

If one were interested in treating $C_\ell^{DD}$ in the data vector then it would be necessary to handle the high $k$ sensitivity since $P_{ee}$ is very uncertain at high $k$.  Within our framework a natural way to do this is through the use of ``counterterms'' (see e.g.\ \cite{DeRose26b} for the same idea in the weak lensing case).  Specifically imagine $P_{ee}=P_{\rm model}+\delta P$ where $\delta P$ is non-zero only for $k>\Lambda$.  Then
\begin{align}
    C_\ell^{DD} &= \bar{n}_{e,0}^2 \int_0^{\chi_f} d\chi\ \frac{(1+z)^2}{\chi^2}\, P_{ee}(k=[\ell+1/2]/\chi,z) \\
    &= C_\ell^{\rm model} + \frac{\bar{n}_{e,0}^2}{\ell+1/2} \int_{\Lambda}^{\infty} dk\ (1+z)^2\ \delta P(k)
    \qquad \left(\ell<\Lambda\,\chi_f-\tfrac{1}{2}\right) \\
    &= C_\ell^{\rm model} + \frac{\mathrm{const}}{\ell+1/2}
\end{align}
where we have assumed $f_e(z)=f_{e,0}$ and are implicitly only considering $\ell<\Lambda\chi_f\gg 1$, as $\ell>\Lambda\chi_f$ are dominated by noise. For large $\Lambda$ we can approximate $1+z\approx 1$ to provide an estimate for the constant which can be used as a prior in any fit to data.

\bibliographystyle{JHEP}
\bibliography{main}

\end{document}